# Explainability and Transferability of Machine Learning Models for Predicting the Glass Transition Temperature of Polymers

Agrim Babbar[1], Sriram Ragunathan[2], Debirupa Mitra[1*], Arnab Dutta[1*], and Tarak. K Patra[2*]

[1]Department of Chemical Engineering, Birla Institute of Technology and Science Pilani, Hyderabad Campus, Shameerpet, Hyderabad, Telangana 500078, India
[2]Center for Atomistic Modeling and Materials Design, Department of Chemical Engineering, Indian Institute of Technology Madras, Chennai TN 600036, India

**Abstract**

Machine learning (ML) offers promising tools to develop surrogate models for polymer's structure-property relations. Surrogate models can be built upon existing polymer data and are useful for rapidly predicting the properties of unknown polymers. The accuracy of such ML models appears to depend on the feature space representation of polymers, the range of training data, and learning algorithms. Here, we establish connections between these factors for predicting the glass transition temperature ($T_g$) of polymers. Our analysis suggests linear models with a smaller number of fitting parameters are as accurate as nonlinear models with a large number of hidden and unexplainable parameters. Also, the performance of a monomer topology-based ML model is found to be qualitatively identical to that of a physicochemical descriptor-based ML model. We find that the transferability of ML models enhances as the property range of the training data increases. Moreover, we establish new $T_g$ – polymer chemistry correlations via ML. Our work illustrates how ML can advance the fundamental understanding of polymer structure-property correlations.

*Corresponding Authors,
E-mail: AD: arnabdutta@hyderabad.bits-pilani.ac.in
       DM: debirupa@hyderabad.bits-pilani.ac.in
       TKP: tpatra@iitm.ac.in

# I. Introduction

In recent years, there has been significant interest and progress in developing machine learning (ML) models for predicting the properties of materials, including polymers.[1–18] These models use large datasets of materials properties and their corresponding features, such as composition, crystal structure, bonding patterns, and other chemical descriptors to train algorithms and use them to predict the properties of unknown materials. However, polymer science encompasses a vast range of chemical space with diverse properties, compositions, and structures. Developing ML models that can effectively handle this complexity and generalize across different polymer classes remains challenging. This requires a deeper understanding of polymer chemistry, topology, feature space representation, selection of data points, and ML algorithms. Such generalizable and transferable ML models have a huge implication in polymer design. They are often used to predict the properties of polymers within a design cycle to fasten the discovery of new polymers.[19,20] These design algorithms try to direct the search toward the unknown region of the physicochemical space of a polymer. However, ML models are inherently interpolative, and their ability to search for extremal candidates is unknown. The primary bottleneck in developing transferable and generalizable models is the quality and quantity of available data, as measuring the properties using physics-based methods for a large number of candidate structures is time and resource intensive. Hence, building transferable ML models with small data remains an attractive goal of polymer informatics.

Moreover, developing an ML model involves two main steps. First is the feature engineering wherein the most relevant features, which are also known as fingerprints, for the model are decided. Molecular fingerprints are broadly two types - topological fingerprints that encode the connectivity of atoms in a molecule, and physicochemical fingerprints that incorporate information such as polarity, hydrophobicity, molecular weight, etc. It is not clear which fingerprint strategy to be used for building an ML model. How does the performance of an ML model compare with these two kinds of fingerprints? The second step of ML model development is the algorithm selection wherein an appropriate learning algorithm for a problem is chosen. This may involve selecting from a range of algorithms such as linear regression, decision trees, neural networks, etc. Again, there is no specific guideline for algorithm selection. Typically, one builds different models to identify the best one for a given problem.

It is also desirable that a structure-property model provides the underlying principles and mechanisms governing polymer behavior. Developing ML models that are explainable and provide insights into the underlying physical and chemical factors of a polymer's behaviors is an ongoing challenge. Algorithms such as decision trees or linear regression, can provide natural explanations for their predictions. These models have a transparent structure that humans can easily understand. However, many ML models, such as artificial neural networks, are often considered as black boxes. It is difficult to explain the prediction and all the internal parameters of a neural network. For such complex models, SHAP (SHapley Additive exPlanations) analysis[21] can be used to identify the role of an individual input feature in determining a specific property of polymers. Towards this end, formulating a feature space depending on the target correlations is a challenging task.

Here, our objectives are to address above challenges, understand how an ML model's performance is connected to fingerprinting and learning algorithm, and study its transferability. In addition, we aim to identify correlations of a polymer property with its physicochemical descriptors including hydrogen bonds, hydrophobicity, hybridization of carbon atoms, monomer topology (cyclic or linear) via machine learning. As a representative case, we choose the glass transition temperature ($T_g$) of a polymer as a target property. There have been many recent attempts to develop ML models for predicting the glass transition temperatures of polymers.[22–31] Motivated by these previous studies, here, our objective is to study the explainability and transferability of three different ML models that predict $T_g$ of a polymer based on two different molecular fingerprints. We have extracted a set of physicochemical fingerprints of monomer moieties from the RDKit[32] open-source software and used these as descriptors to develop ridge regression (RR) and artificial neural network (ANN) models. We have also obtained a topological fingerprint of monomers via one-hot encoding (OHE) of their SMILE (Simplified molecular-input line-entry system)[33] representations and used these to develop a convolutional neural network (CNN) model. Using these three ML models for predicting the $T_g$ of polymers, we have assessed the reliability of an ML model outside the training range and extracted physical insights that are useful in establishing new correlations between $T_g$ and monomer chemistry. Moreover, the number of fitting parameters in an ML model is an important factor as it impacts the model's complexity, memory requirements, and model's ability to learn from data. Models with too few parameters may struggle to capture complex patterns, while models with too many parameters may often overfit and generalize poorly to new data. We choose models with varying range of adjustable parameters, and level of abstraction and establish their connection with the model's performance. We expect the current study will serve as an important reference to develop more transferable and accurate ML models and advance the current understanding of ML models' explainability with respect to polymer chemistry and its impact on the glass transition.

## II. Polymer Data

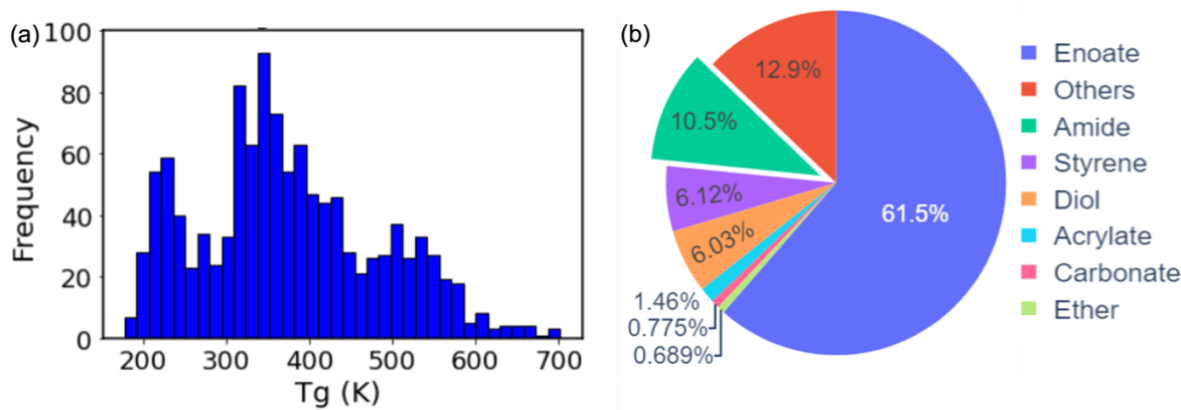

*Figure 1: Polymer data. The distribution of glass transition temperature in the data set is shown in (a). The representation of different chemical moieties in the data set is shown in (b). There are a total of 1162 data, they are collected from the PolyInfo database.*

We have mined glass transition temperature ($T_g$) data of different classes of polymers from the PolyInfo database[34] as shown in Figure 1. They correspond to experimentally measured $T_g$ values of homopolymers. Here, we plan to predict the $T_g$ based on the information on monomer chemistry. We assume that the $T_g$ values are not influenced by the molecular weight, which is usually the case for large molecular-weight polymers. The $T_g$ values in this data set varies from 180K to 700K. It comprises of most commonly used chemical moieties as shown in Figure 1b with a large pool of enolates.

## III. ML Algorithms

The objectives of this study are achieved by deploying three ML algorithms - ridge regression (RR), artificial neural network (ANN), and convolutional neural network (CNN) to predict the $T_g$ value of polymers. Each of these algorithms represent different levels of complexity and extent of explainability. We use physicochemical fingerprinting for RR and ANN models whereas topological fingerprint viz. an OHE of the SMILE representation of the monomer of a homopolymer for the CNN model development. First, we describe the workflows to building these two classes of ML models.

**Physicochemical descriptor-based ML model:** We extract 208 physicochemical descriptors for 1162 monomers from the RDkit package in Python. We perform a Pearson correlation coefficient analysis to understand how different physicochemical descriptors impact the glass transition temperature of the materials. The range of the coefficient value is (-1,1). A coefficient value 0 indicates no correlation between a physicochemical descriptor and the $T_g$. Similarly, a coefficient value of 1 represent strongest correlation between the descriptor and the $T_g$, and a negative coefficient corresponds to inverse correlation. We set a minimum of the modulus of a correlation coefficient is 0.7 for a descriptor to be selected for the model development. Based on this criterion, we choose 113 physicochemical descriptors and use them for ML model development. These physicochemical descriptors are used as the inputs for the RR and ANN models. Figure 2 represents the workflow of the ML model development based on physicochemical descriptors. The RR is a linear regression model that adds a penalty term to the ordinary least squares (OLS) method in order to prevent the overfitting.[35] It is a regularized regression method that shrinks the regression coefficients towards zero, which helps to reduce the impact of the noise in the data and improve the generalization performance of the model. The penalty term in ridge regression is called the L2-norm penalty and it is computed as the sum of the squares of the regression coefficients multiplied by a tuning parameter called the regularization parameter, denoted by λ. Once the model is trained, we analyze its performance on different data set by estimating the coefficient of determination ($R^2$).

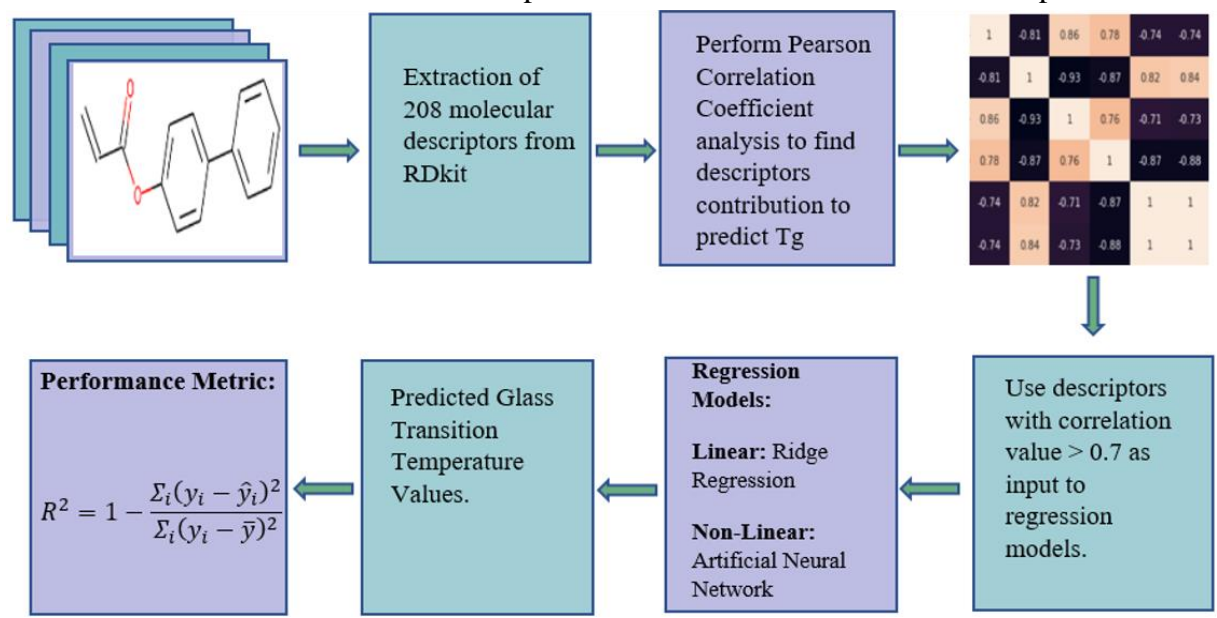

*Figure 2: The workflow for ML model development based on physicochemical descriptors. Monomer SMILES representation is fed to the RDKit and 208 physicochemical descriptors are extracted. Pearson correlation coefficients of all the descriptors are calculated. The descriptors whose Person correlation coefficient are greater that a cut-off value (=0.7) are used as inputs to ML models.*

Similarly, we build an ANN model with the same set of physicochemical descriptors as the input. An ANN attempts to reproduce the brain's logical operation using a collection of neuron-like nodes to process signals.[35,36] It is a nonlinear framework to capture the relationship between input variables (e.g., molecular structure descriptors) and output variables (e.g., material properties). An ANN architecture usually consists of nodes categorized into an input layer, one or more hidden layers, and an output layer. The nodes in the input layer represent the descriptors of the molecule of a material and the output layer contains nodes that represent the properties of the material. Each node in the hidden layers consists of a function that transforms the sum of the weighted input values (consisting of all output values from the previous layer) into an output value that is passed to the next layer. We use sigmoidal function for the activation of signal in all the nodes in the hidden layers. The layers in an ANN are arranged in the form of a directed acyclic graph, with each node of each layer receiving input from all nodes of the prior layer but not from nodes in the same or later layers. Every layer has a bias node that does not receive any input from its previous layer, and it provides a constant bias to all of the nodes in its succeeding layer. The optimal numbers of hidden layers and nodes in a hidden layer have no universal values but are selected so as to maximize the efficiency and accuracy of an ANN model. The network is trained prior to use by optimizing all of the weights in the network using a back-propagation algorithm.[37] The backpropagation algorithm calculates the gradient of an error function, which is the square of the difference between the target and actual outputs with respect to all of the weights in the network. It then uses an optimization method such as gradient descent to update the weights in order to minimize the loss function. The optimal ANN architecture in this study consists of 2 hidden layers. The first and second hidden layers consist of 20 and 30 nodes respectively. We use the rectified linear activation function (RELU) to activate the inputs of all nodes in the hidden layers. The last layer consists of one node that represent the output i.e., $T_g$ of a polymer. We use the Adam optimizer with a learning rate of 0.05 and a batch size of 100 for training the ANN. The loss function is defined as the mean absolute error between the predicted and actual $T_g$ values of polymers.

**Topological descriptor-based ML model**: Topological descriptors capture the connectivity between atoms and the arrangement of bonds in a monomer without considering the spatial orientation or the 3D structure of a monomer in a polymeric material. In this approach, we use an OHE algorithm to convert the SMILES string of a monomer into a binary matrix. Subsequently, the binary matrix is converted to a binary image, as shown schematically in Figure 3, which serves as the input to the ML model. A dictionary of all possible SMILES characters in our data set is first created. Since our database is made of 35 unique SMILES characters, the size of the dictionary is 35, which is represented as: ['c', 'n', 'o', 'C', 'N', 'F', '=', 'O', 'H', '%','0',']', '[', 'Li', '(', ')', '1','\ ','2','#','Cl','/','S',' Br', '3', '4', '.', '5', '6', '7', '8', '+','-', '9', 'P']. Using this dictionary, we transform a SMILES string into a binary matrix whose elements are either zero or one. The length of the largest SMILES code in our database and the number of characters in the dictionary represent the number of rows and columns of the matrix, respectively. Hence, the matrix has 35 columns each representing a unique SMILES character. The sequence of columns is according to the sequence of corresponding characters in the dictionary. The rows represent the SMILES characters of a given monomer. An element of the matrix is zero if the labels of the corresponding row and column are different. In case, the row and column labels of an element match, the element is one. This OHE scheme is applied to all the polymers in our data sets, and 1162 binary images are generated. We build the CNN model with these binary images as inputs. The hidden features from these binary images are extracted using two convolutional 2D layers with 64 and 32 filters, respectively. The window size of the two filters is (5,5) and (3,3), respectively. The signal in both the layers is activated by the RELU. The convolutional-2D layers are followed by a Max-Pooling layer with a window size of (3,3). To obtain the glass transition temperature of polymers as the output from the CNN model, the max-pool layer is connected to dense layers. We chose two dense layers consisting of 32 and 10 nodes, respectively. The last dense layer is connected to the output layer node. The loss function is defined as the mean absolute error between the predicted and actual $T_g$ of polymers. We use the Adam optimizer with a learning rate of 0.03 and batch size of 64. A schematic representation of the CNN architecture is shown in Figure 3. The hyperparameters of the CNN model are tuned in order to improve its performance.

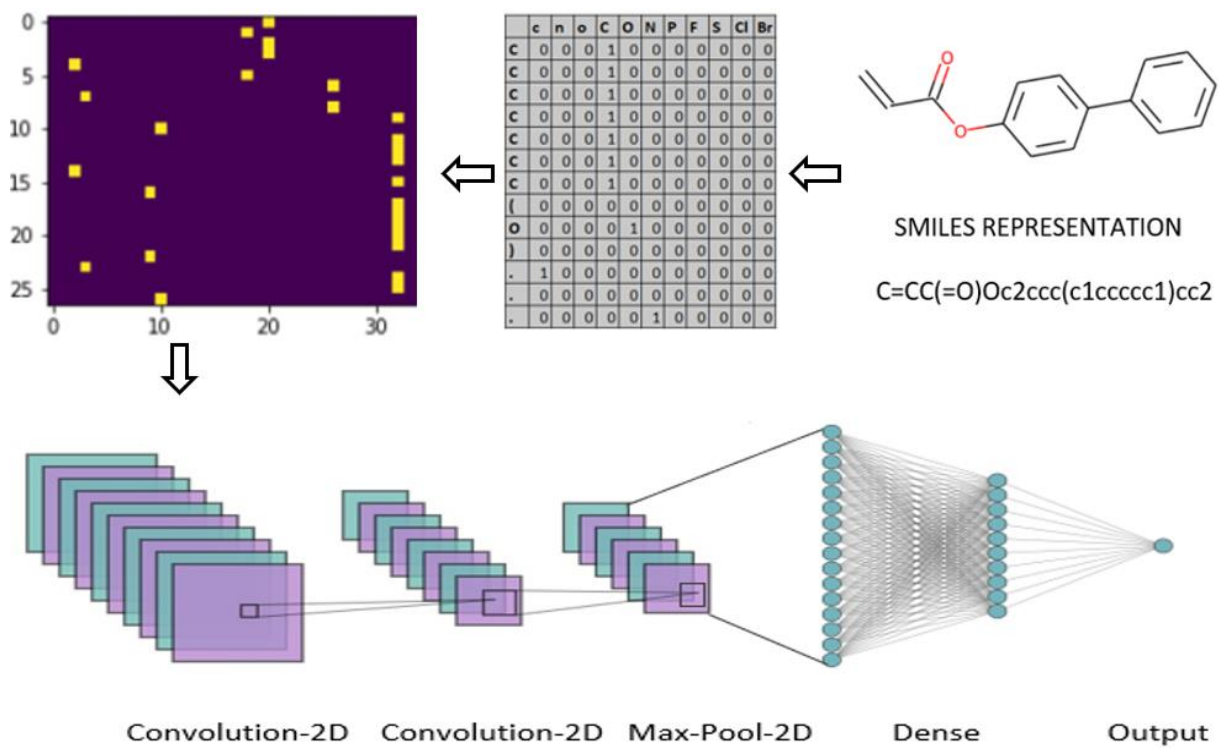

*Figure 3: The workflow of a CNN model based on a topological descriptor. An adjacency matrix is made from the SMILES representation of a monomer. The matrix is converted to a binary image which serves as the input to the CNN. The CNN consists of convolution, pooling and dense layers arranged successively in between the input and output layers.*

## IV. Results and Discussion

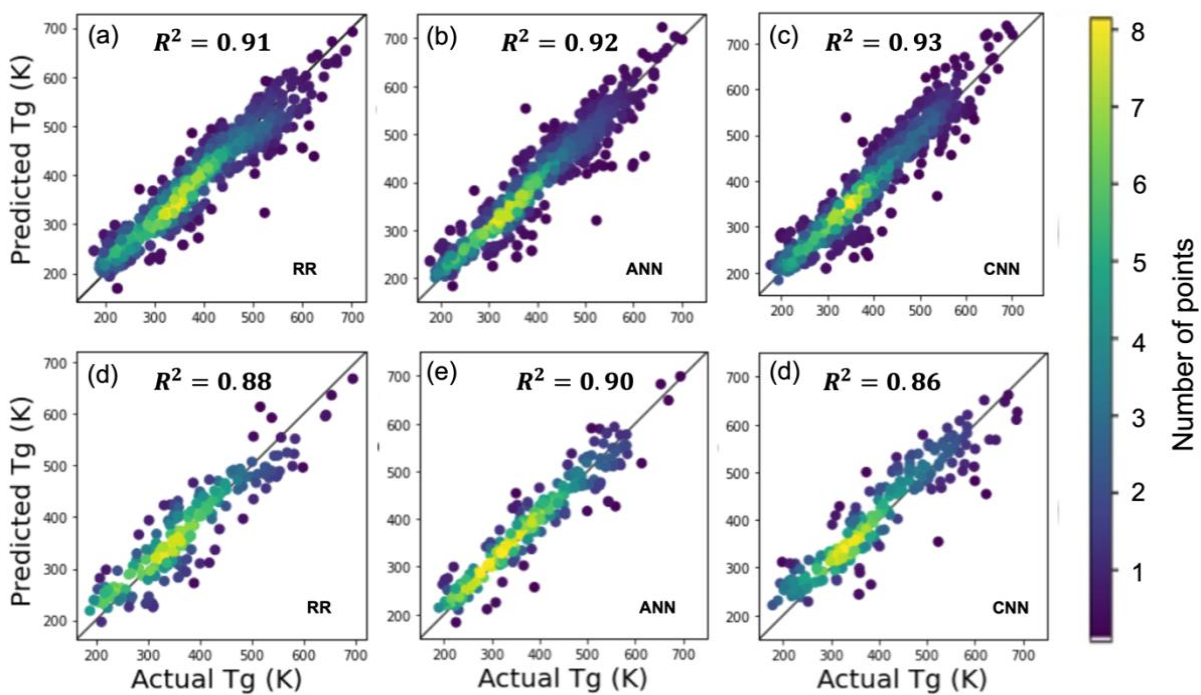

*Figure 4: The performance of ML models. The ML predicted Tg is plotted against the actual Tg of a polymer for all three cases. The top and bottom rows correspond to training and test data sets, respectively. The first, second and third columns correspond to RR, ANN and CNN models, respectively.*

**Performance of ML models:** We begin by training ML models with 80% of data and test models' performance for the remaining 20% of the data. Figure 4 shows the performance of all three ML models. The training and testing of the models are done based on the same data points. We compare the performance of the models bases on their coefficient of determinations ($R^2$). The training and test $R^2$ scores for RR, ANN, and CNN are comparable, as indicated in Figure 4. It is interesting to see that the CNN model also results in a similar level of accuracy as of RR and ANN. The $R^2$ of the CNN model on training is found to be only marginally higher than that of the physicochemical descriptor-based models. However, the $R^2$ is found to be the lowest for the CNN model with respect to the test data set. Besides $R^2$, we also estimate the Akaike Information Criteria (AIC) score for the ML models. The AIC score considers both performance as well as number of parameters used in an ML model. It is to be noted that the absolute value of the AIC is not significant; however, a lower value of AIC score indicates a relatively better model when several models are compared. The AIC score is defined as $AIC = n * \log\left(\frac{\sum(y_i - \hat{y}_i)^2}{n}\right) + 2 * p$, where n and p denote the number of samples and the number of parameters to be estimated by the model, respectively. Here, $y_i$ represents the true value of the property and $\hat{y}_i$ represents the predicted property value from the ML model. As shown in Table 1, the number of parameters is maximum for the CNN model and minimum for the RR model.

Consequently, the AIC score is found to be the lowest for the RR model and highest for the CNN model. It clearly suggests that a complex model with higher number of parameters does not always result in more accurate predictions. Thus, it can be inferred that a much simpler regression model is able to predict $T_g$ of a polymer with about the same accuracy as compared to more complicated nonlinear ANN or CNN models.

| ML Model | Number of Adjustable Parameters | AIC Score |
|---|---|---|
| RR | 114 | 8437.638 |
| ANN | 2941 | 12997.102 |
| CNN | 564245 | 1136663.631 |

*Table 1: Fitting parameters and AIC scores for all the ML models are shown in this table. The number of adjustable parameters correspond to the best performance of a model.*

**Transferability of ML models:** Next, we want to examine the transferability of ML models within the context of polymer $T_g$ prediction. We systematically build models with a chosen range of $T_g$ and analyze their performance outside the range of training data. Each time, we build a model, we adjust the associated hyperparameters to achieve its best performance. The mean $T_g$ in our dataset is 374.30K. The training data are collected around the mean $T_g$. First, we consider a case where the training data points are within ± 5% of the mean $T_g$, and the remaining points are used for testing. We repeat this exercise for six more cases wherein the training datasets consist of points that are ± 10%, 15%, 20%, 25%, 30% and 50% of the mean Tg. The absolute error in model's prediction for both the training and test data sets is shown in Figure 5 for three representative cases. All the three models make reasonably good prediction for the data points that are within the training range as well as from the nearby region of the training data range. Clearly, the error in prediction increases as we go far away from the training data range. As shown in Figure 6, the quality of prediction improves as the range of training data expands. However, both the ANN and RR models have reasonably higher $R^2$ than that of the CNN for smaller range of training data. The $R^2$ of the CNN model becomes comparable to that of the ANN or RR when the training data size is about 700.

**Explainability of ML models:** We now focus on understanding the correlations between the physicochemical descriptors and $T_g$ of a polymer. We perform the SHAP[21] analysis on the ANN model to find out the top 20 features, which are shown in Figure 7a. In addition, we also rank the physicochemical descriptors according to their slope in the RR model, and report in Figure 7b. On comparing these two figures, we observe that the top features obtained from the ANN model are identical to those obtained from the RR model. Table 2 gives a brief description of the most important physicochemical descriptors that determine the $T_g$ of a polymer.

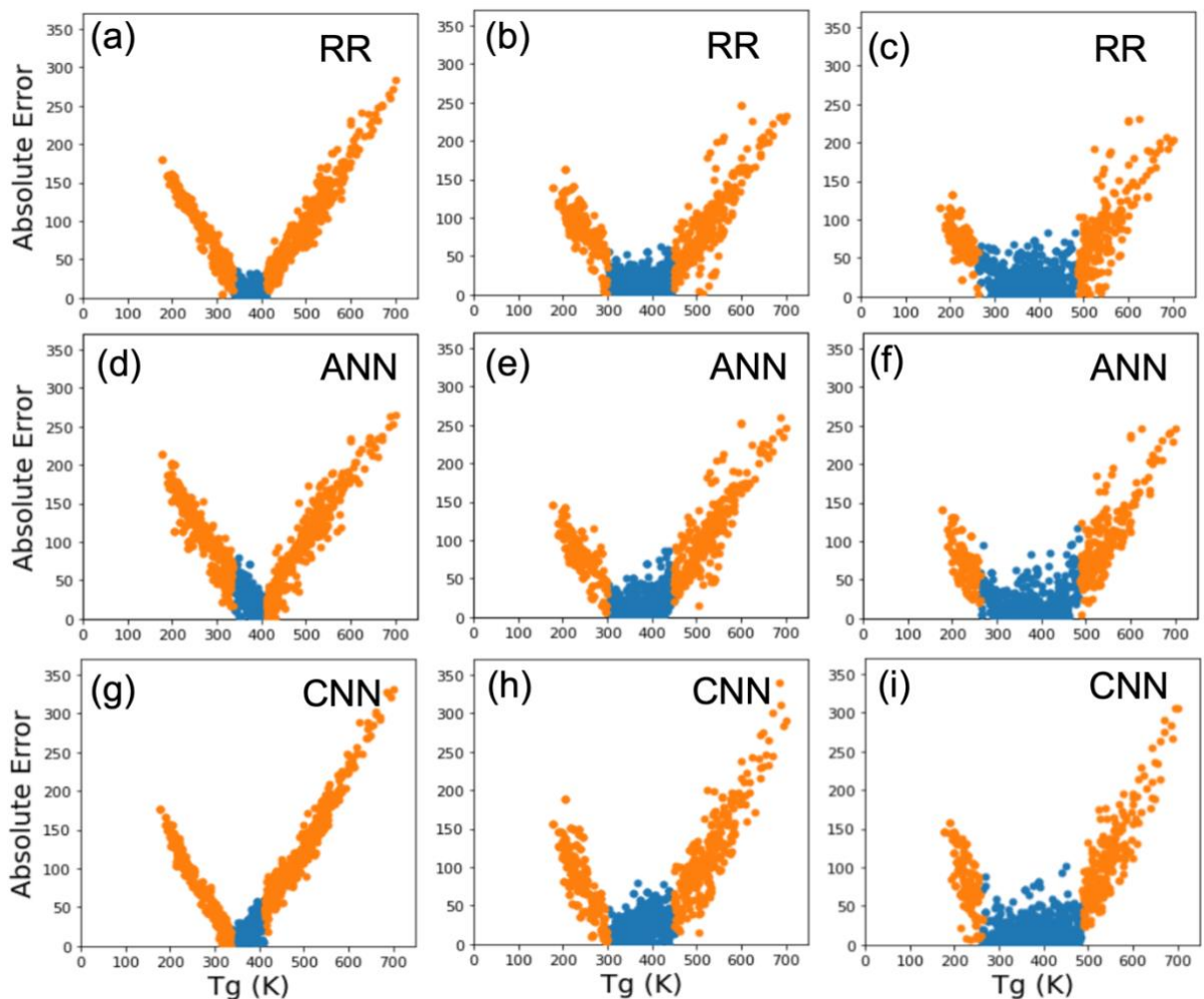

*Figure 5: Transferability of ML models. The absolute error in ML model's prediction is plotted as a function of the actual $T_g$ of a polymer. Blue points are training data and the orange points are test data. The first, second and third columns correspond to 10%, 20% and 30% of data from the database, respectively.*

Our analysis suggests that the fraction of carbon atoms that are in *sp3* hybridization state (*FractionCSP3*) and the number of rotatable bonds (*NumRotatableBonds*) in a monomer are the most important features of a polymer that determine its $T_g$. Additionally, both these physicochemical descriptors have an inverse correlation with the $T_g$. A high fraction of C atoms in the *sp3* hybridization state indicates a low fraction of C atoms in *sp2* (double bond) or *sp* (triple bond) hybridization states. Since the presence of double or triple bonds is expected to increase the stiffness of any polymer by hindering bond rotation, a low fraction of these will increase the stiffness of any polymer by hindering bond rotation, a low fraction of these will result in lower $T_g$. Similarly, higher the number of rotatable bonds, higher will be the chain mobility thereby decreasing the $T_g$ of any polymer. The two most important physicochemical descriptors that directly correlate to $T_g$ of a polymer are the number of NH and OH groups (*NHOH Count*) and the total number of rings (*RingCount*) present in a monomer. Considering *NHOHCount*, it is well-known that both NH and OH functional groups participate in inter and intra molecular hydrogen bonding. Increased hydrogen bonding within and between chains lead to an increment in $T_g$. Thus, higher the number of NH and OH groups, higher will be the $T_g$ of a polymer. Similarly, increase in the number of H donors also results in increased $T_g$ values. The presence of rings is also associated with higher $T_g$ values. Any molecule that is cyclic or has rings has substantially lower configurational entropy as compared to a linear molecule with the same number of carbon atoms resulting in a higher value of $T_g$.[38]

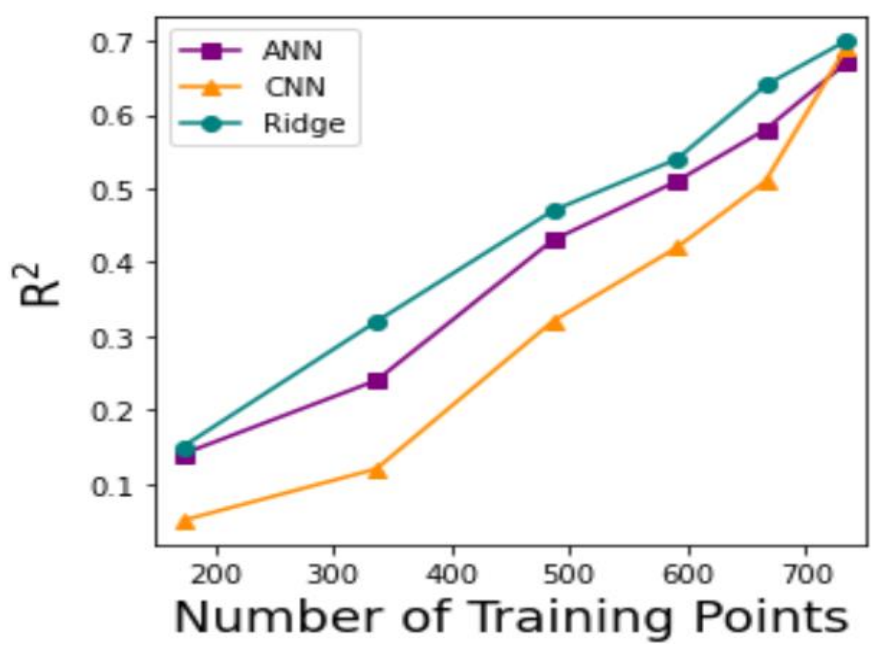

*Figure 6: Transferability of an ML model. The coefficient of determination $R^2$ for test data set is plotted against the number of training data for three models. Training and test data are from different range of $T_g$.*

Another descriptor that is also associated with configurational entropy is the *Kappa2 index*. The Kappa indices, formulated by Hall and Kier, are indicators of molecular shape.[39] The *Kappa2 index* in particular encodes information about spatial arrangement of atoms in a molecule along with relative size contribution of the atoms with respect to the *sp3*-hybridized C atom. As per this index, molecules with higher configurational entropy have higher *Kappa2* values. Our analysis also shows that molecules with higher *Kappa2* values is more likely to result in a lower $T_g$. We also observe that the combined effect of atomic contributions towards the monomer surface area (as measured by the van der Waal's surface area, *VSA*) and electrotopological state of atoms in the monomer (as measured by the *E-state index*),[40] or the hydrophobicity of a monomer (as measured by *log P* i.e. octanol-water partition coefficient values), or polarizability (or molecular refractivity) of the monomer are important factors that determines the $T_g$ of a polymer.

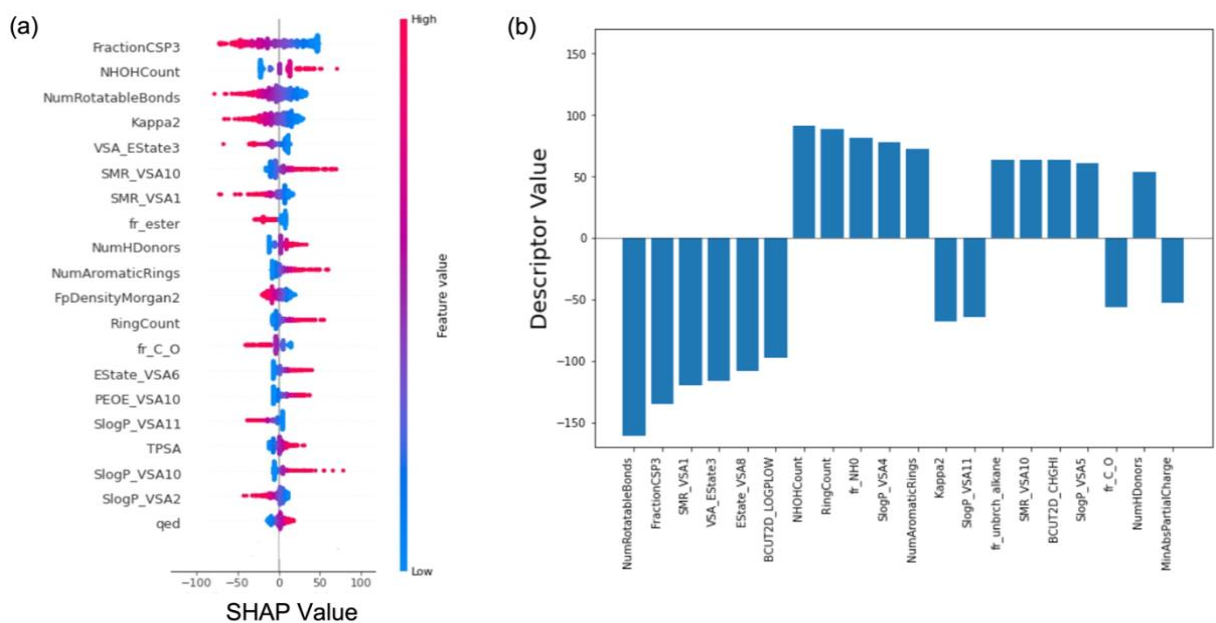

*Figure 7: Impact of chemical descriptors on the polymer Tg. The SHAP scores for top 20 descriptors of the ANN model are shown in (a). The weights of the top 20 chemical features in the RR model are shown in (b).*

The *VSA_Estate* descriptors represent a combined *E-state* value of atoms whose contribution towards *VSA* lies in a particular range. For example, *VSA_Estate3* is the sum of the *E-state* values of all atoms in the molecule (i.e., the monomer) with *VSA* value between 5.00 and 5.41 Å$^2$. In a similar fashion, the *SMR_VSA* and *SlogP_VSA* descriptors are the total *VSA* of atoms that have a particular range of molecular refractivity (MR) and *log P*, respectively (*cf.* Table 2). Both *MR* and *log P* values are calculated for every atom type based on a method developed by Wildman and Crippen.[41] It is interesting to note that even a particular type of these combined descriptors can impact the $T_g$ of a polymer very differently. For instance, the *SMR_VSA1* affects the Tg negatively, while the *SMR_VSA10* affects it positively. If the total *VSA* of atoms having *MR* between -∞ to 1.29 (*SMR_VSA1*) is high, it is likely to result in a lower $T_g$, whereas if the total *VSA* of atoms having *MR* between 4 to ∞ (*SMR_VSA10*), it is likely to result in a higher $T_g$. Some of the atom types which contribute to *SMR_VSA1* are aliphatic amine *N*, protonated amine *N*, protonated aromatic *N*, aromatic *O* and carbonyl aliphatic *O*. Some of the atom types which contribute to *SMR_VSA10* are *C* doubly bonded with a heteroatom, aromatic *C* with a heteroatom neighbor, aromatic bridgehead *C*, and aromatic C=C. Similar observation is noted for some of the other combined descriptors. For instance, *SlogP_VSA11* and *SlogP_VSA5* affect $T_g$ differently (Figure 7b). If the total *VSA* of atoms having *logP* between 0.5 to 0.6 (*SlogP_VSA11*), it is likely to result in a lower *T*$_g$, whereas if the total *VSA* of atoms having *logP* between 0.10 to 0.15 (*SlogP_VSA5*), it is likely to result in a higher *T*$_g$. The only atom type that contributes to *SlogP_VSA11* is a 4° aromatic *C* with *O* as the neighboring atom.[41] Some of the atom types which contribute to *SlogP_VSA5* are 1° or 2° aliphatic C, 3° aromatic C, 4° aromatic C with C as the neighboring atom, hydrocarbon H, and carbonyl aromatic O.

| Descriptor | Definition |
| --- | --- |
| FractionCSP3 | The fraction of sp3-hybridized carbon atoms in a monomer |
| NHOH Count | The number of N-H and O-H groups in a monomer |
| NumRotatableBonds | The number of rotatable bonds in a monomer |
| Kappa2 | Characterizes the topological structure of a monomer |
| VSA_EState3 | The group contribution of electrotopological states of all atoms in a monomer whose Van der Waal's surface area lies between 5.00 to 5.41 Å$^2$ |
| SMR_VSA10 | The group contribution of Van der Waal's surface area of all atoms in a monomer whose molecular refractivity is greater than 4 |
| SMR_VSA1 | Represents the group contribution of Van der Waal's surface area of all atoms in a chemical compound whose molecular refractivity is lesser than 1.29 |
| fr_ester | Quantifies the number of ester functional groups in a chemical compound |
| NumHDonors | Quantifies the number of hydrogen bond donor groups in a chemical compound |
| NumAromaticRings | Quantifies the number of aromatic rings present in a chemical compound |
| FpDensityMorgan2 | Represents the density of Morgan fingerprints with a radius of 2 Å |
| RingCount | Quantifies the number of rings present in a monomer |
| fr_C_O | The number of carbon-oxygen (C-O) single bonds in a monomer |
| Estate_VSA6 | Represents the group contribution of Van der Waal's surface area of all atoms in a chemical compound whose Estate indices lie between 1.54 and 1.81 |
| PEOE_VSA10 | Represents the group contribution of Van der Waal's surface area of all atoms in a chemical compound whose partial charges (calculated using the PEOE (Partial Equalization of Orbital Electronegativities) method) lie between 0.10 to 0.15 |
| SlogP_VSA11 | Represents the group contribution of Van der Waal's surface area of all atoms in a chemical compound whose log P values lie between 0.5 and 0.6 |
| TPSA | Quantifies the approximate polar surface area (Å$^2$) of a monomer |
| SlogP_VSA10 | Represents the group contribution of the Van der Waal's surface area of all atoms in a monomer whose log P values lie between 0.4 and 0.5 |
| SlogP_VSA2 | Represents the group contribution of Van der Waal's surface area of all atoms in a chemical compound whose log P values lie between -0.4 and -0.2 |
| Qed | Single numerical value that indicates how likely a molecule is to have drug-like properties |
| Estate_VSA8 | Represents the group contribution of Van der Waal's surface area of all atoms in a monomer whose Estate indices lie between 2.05 and 4.69 |
| BCUT2D_LOGPLOW | Quantifies the contribution of different atom pairs to measure a monomer's lipophilicity |
| fr_NHO | Quantifies the number of N-H-O linkages in a monomer |
| SlogP_VSA4 | Represents the group contribution of Van der Waal's surface area of all atoms in a monomer whose log P values lie between 0 and 0.1 |
| fr_unbrch_alkane | Quantifies the number of unbranched alkanes in a monomer |
| BCUT2D_CHGHI | Quantifies the contribution of different atom pairs to the calculated partial charge of a monomer |
| SlogP_VSA5 | Represents the group contribution of Van der Waal's surface area of all atoms in monomer whose log P values lie between 0.1 and 0.15 |
| MinAbsPartialCharge | Quantifies the minimum absolute value of the partial charges of atoms in a monomer |

*Table 2: Description of physicochemical descriptors of a polymer that are identified via ML to have significant impact on its T$_g$*

We note that some of these properties capture information at the monomer level like the total number of rings or the total number of NH and OH groups in a monomer, while others do so at the atomic level such as the VSA-associated combined descriptors.

## Conclusions

The chemical space of a polymer is so vast that it is almost impossible to screen the entire space via physics-based measurements. Machine learning is a potential avenue that promises to tackle this problem. A large numbers of ML models are reported in recent times that make rapid prediction of polymer properties. A wide variation of feature space representation and learning algorithms are proposed to build ML models for polymers. Here, we study the complexity, explainability and transferability of these models for a representative case of glass transition temperature of polymers. We choose three commonly used algorithms viz. RR, ANN, and CNN for predicting glass transition temperature of a polymer. We derive suitable feature space representations that are compatible with these learning algorithms. The RR and ANN models are built based on physicochemical descriptors of the monomer of a polymer. Apparently, the performance of the RR and the ANN are comparable. The RR model provides weightage of different physicochemical descriptors that can be directly used to decipher the impact of these descriptors on glass transition temperature of polymers. On the contrary, the ANN model being nonlinear in nature is not directly explainable. To extract meaningful insights from the ANN model, we have implemented SHAP analysis to correlate $T_g$ prediction with polymer chemistry. Although the interworking of the ANN model is less transparent, we observe that it does verify the descriptor-property correlations that are established via a linear model such as RR. Further, we use topological fingerprint of monomers to develop a CNN model. The performance of the CNN model is found to be comparable to RR and ANN models, although the complexity and computational costs are significantly higher than the other two. In terms of transferability, the CNN model is inferior than the other two in the low data limit, and it is observed to improve with higher data and become comparable to that of chemical fingerprints-based models (i.e., RR and ANN). Overall, we infer that a linear model such as RR can not only make reasonably accurate prediction of glass transition temperature of polymers but also they are more explainable as compared to nonlinear models. Contrary to the perception that more interpretable models may scarifies predictive performance, here we observe that the RR performance is relatively better or comparable to the nonlinear models like ANN and CNN. Also, our analysis suggests that a complex model with large number of parameters struggles to learn in the low data limit and often results in poor predictions in outside the training range. The explainable machine learning technique, implemented in this study, helps us to unveil the correlation between $T_g$ of a polymer with its physicochemical descriptors such as carbon atoms hybridization, hydrophobicity, polarity, and electrotopological state. We envision that the insights obtained from this work will have important implications in more accurate and explainable ML model development for polymer property prediction.

## Acknowledgment

The work is made possible by financial support from the SERB, DST, Gov of India through a core research grant (CRG/2022/006926)) and the National Supercomputing Mission's research grant (DST/NSM/R&D_HPC_Applications/2021/40). This research used resources of the Argonne Leadership Computing Facility and Center for Nanoscience Materials, which are DOE Office of Science User Facilities supported under the Contract No. DE-AC02-06CH11357.